\documentclass[aps, prd, onecolumn, showpacs, showkeys, superscriptaddress, preprintnumbers, floatfix, reprint]{revtex4}
\usepackage{slashed}
\usepackage{multirow}
\usepackage{graphicx}
\usepackage{amsmath}
\usepackage{bm}
\usepackage{yhmath}
\usepackage{hyperref}
\usepackage{subfigure}
\usepackage{color}
\usepackage{cases}
\usepackage{subfigure}
\usepackage{dcolumn, booktabs, bm}
\usepackage{amsfonts, amssymb, stmaryrd, latexsym, amsmath}
\usepackage{textcomp}

\usepackage{array}
\newcounter{RomanNumber}

\newcommand{\lyxmathsym}[1]{\ifmmode\begingroup\def\b@ld{bold}
	\text{\ifx\math@version\b@ld\bfseries\fi#1}\endgroup\else#1\fi}

\allowdisplaybreaks
\UseRawInputEncoding
\begin{document}
	\title{Chiral corrections to the masses of the doubly heavy baryons}
	\author{Hao-Ze Tong}\email{hztong@stumail.nwu.edu.cn}\affiliation{School of Physics, Northwest University, Xian 710127, China}
	\author{Hao-Song Li}\email{haosongli@nwu.edu.cn}\affiliation{School of Physics, Northwest University, Xian 710127, China}\affiliation{Institute of Modern Physics, Northwest University, Xian 710127, China}\affiliation{Shaanxi Key Laboratory for theoretical Physics Frontiers, Xian 710127, China}\affiliation{Peng Huanwu Center for Fundamental Theory, Xian 710127, China}
	\begin{abstract}
		We study the masses of the doubly bottom baryons and the charmed bottom baryons up to $\mathcal{O}\left(p^{3}\right)$ in heavy baryon chiral perturbation theory. For the low-energy constants, $g_{A}$ was determined in the quark model, and we fit to the lattice QCD data to determine $m_{0}$, $c_{1}$, and $c_{7}$. We show the numerical results up to $\mathcal{O}\left(p^{3}\right)$, and our results are compared with other theoretical results.
	\end{abstract}
	\maketitle
	\thispagestyle{empty}
	\section{Introduction}
	\label{sec1}
	A key prediction of QCD is the existence of baryons with two heavy quarks and one light quark. The SELEX collaboration first found the $\Xi^{+}_{cc}$ baryon with the mass $m_{\Xi^{+}_{cc}}=3519\pm1{\rm MeV}$ \cite{Mattson: 2002vu}. However, this result was not confirmed by other collaborations \cite{Ratti: 2003ez, Aubert: 2006qw, Chistov: 2006zj}. In 2017, the LHCb collaboration found the $\Xi^{++}_{cc}$ baryon with the mass $m_{\Xi^{++}_{cc}}=3621.40\pm0.72\pm0.27\pm0.14{\rm MeV}$ \cite{Aaij: 2017ueg}. So far, people did not find experimental evidences on more doubly heavy baryons.
	
	Nowadays, people have widely studied the masses of the doubly heavy baryons. In the quark model, the authors studied the masses of the doubly heavy baryons in an extended chromomagnetic model \cite{Weng: 2018mmf}, and the masses of the doubly heavy baryons were studied in a hypercentral constituent quark model \cite{Mohajery: 2018qhz, Salehi: 2019kvb}. The authors used different potential models to study the masses of the doubly heavy baryons \cite{Karliner: 2014gca, Karliner: 2018hos, Shi: 2019tji}. In lattice QCD, the authors studied the masses of the doubly charmed baryons \cite{Brown: 2014ena, Bali: 2015lka, Alexandrou: 2017xwd, Bahtiyar: 2020uuj}, and the masses of the doubly bottom baryons and the charmed bottom baryons were also studied \cite{Brown: 2014ena, Lewis: 2008fu, Burch: 2015pka, Mathur: 2018epb, Mohanta: 2019mxo}. Lattice QCD simulations precisely provide the masses of the doubly heavy baryons, but usually give unphysical results, so people must perform chiral extrapolations \cite{Brown: 2014ena, Mathur: 2018epb}. Recently, lattice QCD simulations provide the masses of the doubly heavy baryons near the physical point \cite{Alexandrou: 2017xwd, Bahtiyar: 2020uuj}. Besides, more methods were involved, including Bethe-Salpeter equation \cite{Yu: 2018com, Li: 2019ekr}, contact-interaction models \cite{Gutierrez-Guerrero: 2019uwa, Yin: 2021uom}, and QCD sum rules \cite{Chen: 2017sbg, Wang: 2018lhz}.
	
	Chiral perturbation theory (ChPT) \cite{weinberg: 1978kz} is a low-energy effective field theory of QCD. The power counting scheme \cite{weinberg: 1991um} is used in the meson sector, but breaks down in the baryon sector due to the nonzero baryon mass in the chiral limit \cite{Gasser: 1987rb}. People proposed many schemes to solve this issue, including heavy baryon chiral perturbation theory (HBChPT) \cite{Jenkins: 1990jv}, the infrared (IR) \cite{Becher: 1999he, Kubis: 2000zd} and extended-on-mass-shell (EOMS) schemes \cite{Fuchs: 2003qc}. The principle of the heavy quark effective field theory is used in HBChPT, so people take the nonrelativistic limit and make the expansion in terms of the inverse baryon mass. Later, quenched chiral perturbation theory (QChPT) and partially quenched chiral perturbation theory (PQChPT) were proposed \cite{Bernard: 1993sv, Sharpe: 2000bc, Bernard: 2013kwa}.
	
	For the doubly heavy baryons, the heavy quarks are static, and the light quark governs the chiral dynamics. Nowadays, people have widely studied the chiral corrections to the masses of the doubly heavy baryons. From the heavy quark-diquark symmetry, the authors constructed the Lagrangians to study the masses of the doubly heavy baryons \cite{Hu: 2005gf}, and extend this result to QChPT and PQChPT \cite{Mehen: 2006vv}. The authors studied the masses of the doubly charmed baryons up to $\mathcal{O}\left(p^{4}\right)$ in HBChPT \cite{Sun: 2014aya}. In the EOMS scheme, the masses of the doubly charmed baryons were studied up to $\mathcal{O}\left(p^{3}\right)$ \cite{Sun: 2016wzh}, and this result was extended up to $\mathcal{O}\left(p^{4}\right)$ \cite{Yao: 2018ifh}. In Refs. \cite{Sun: 2014aya, Sun: 2016wzh, Yao: 2018ifh}, the authors showed the numerical results up to $\mathcal{O}\left(p^{3}\right)$.
	
	Similar to the doubly charmed baryons, the chiral corrections to the masses of the doubly bottom baryons and the charmed bottom baryons are still worth exploring. We study the masses of the doubly bottom baryons and the charmed bottom baryons up to $\mathcal{O}\left(p^{3}\right)$ in HBChPT. We show the numerical results up to $\mathcal{O}\left(p^{3}\right)$.
	
	Our paper is organized as follows. We introduce the Lagrangians of the doubly bottom baryons in Sec. \ref{sec2}. We derive the mass formulas of the doubly bottom baryons in Sec. \ref{sec3}. We show the numerical results for the masses of the doubly bottom baryons and the charmed bottom baryons in Sec. \ref{sec4}. Summary is given in Sec. \ref{sec5}.
	\section{The Lagrangians of the doubly bottom baryons}
	\label{sec2}
	In ChPT, the Lagrangians for the Goldstone bosons and the doubly bottom baryons were constructed in Refs. \cite{Sun: 2014aya, Qin: 2019hgk}. The lowest order Lagrangian reads
	\begin{equation}
	\mathcal{L}^{\left(1\right)}_{MB}=\bar{\psi}\left(i\slashed{D}-m_{0}+\frac{g_{A}}{2}\slashed{u}\gamma^{5}\right)\psi,
	\end{equation}
	where $m_{0}$ is the mass of the doubly bottom baryons in the chiral limit, and $g_{A}$ is the axial coupling constant. The Goldstone boson field reads
	\begin{equation}
	\phi=
	\left(
	\begin{array}{ccc}
	\pi^{0}+\frac{\sqrt{3}}{3}\eta&\sqrt{2}\pi^{+}&\sqrt{2}K^{+}\\
	\sqrt{2}\pi^{-}&-\pi^{0}+\frac{\sqrt{3}}{3}\eta&\sqrt{2}K^{0}\\
	\sqrt{2}K^{-}&\sqrt{2}\bar{K}^{0}&-\frac{2 \sqrt{3}}{3}\eta\\
	\end{array}
	\right).
	\end{equation}
	The doubly bottom baryon field reads
	\begin{equation}
	\psi=
	\left(
	\begin{array}{c}
	\Xi^{0}_{bb}\\
	\Xi^{-}_{bb}\\
	\Omega^{-}_{bb}\\
	\end{array}
	\right).
	\end{equation}
	For the building blocks,
	\begin{gather}
	D_{\mu}=\partial_{\mu}+\Gamma_{\mu},\\
	\Gamma_{\mu}=\frac{1}{2}\left(u^{\dagger}\partial_{\mu}u+u\partial_{\mu}u^{\dagger}\right),\\
	u_{\mu}=\frac{i}{2}\left(u^{\dagger}\partial_{\mu}u-u\partial_{\mu}u^{\dagger}\right),\\
	u={\rm exp}\left(\frac{i\phi}{2F_{0}}\right),
	\end{gather}
	where $F_{0}$ is the decay constant of the Goldstone bosons in the chiral limit. We use the experimental values of $F_{P}$ \cite{ParticleDataGroup: 2020ssz}, $F_{\pi}=92{\rm MeV}$, $F_{K}=113{\rm MeV}$, $F_{\eta}=116{\rm MeV}$. The $\mathcal{O}\left(p^{2}\right)$ and $\mathcal{O}\left(p^{3}\right)$ Lagrangians read
	\begin{gather}
	\mathcal{L}^{\left(2\right)}_{MB}=\bar{\psi}\left[c_{1}\left\langle\chi_{+}\right\rangle+c_{2}u^{2}+c_{3}\left\langle u^{2}\right\rangle+\left(c_{4}\left\{u^{\mu}, u^{\nu}\right\}D_{\mu\nu}+{\rm H.c.}\right)+\left(c_{5}\left\langle u^{\mu}u^{\nu}\right\rangle D_{\mu\nu}+{\rm H.c.}\right)+ic_{6}\left[u^{\mu}, u^{\nu}\right]\sigma_{\mu\nu}+c_{7}\tilde{\chi}_{+}\right]\psi,\\
	\mathcal{L}^{\left(3\right)}_{MB}=\bar{\psi}\left(d_{1}\slashed{u}\gamma^{5}\left\langle \chi_{+}\right\rangle+d_{2}\left\{u^{\mu}, \tilde{\chi}_{+}\right\}\gamma_{\mu}\gamma^{5}+d_{3}\left\langle u^{\mu}\tilde{\chi}_{+}\right\rangle\gamma_{\mu}\gamma^{5}+\cdots\right)\psi.
	\end{gather}
	where $c_{i}\left(i=1, \cdots, 7\right)$ and $d_{j}\left(j=1, \cdots, 3\right)$ are the coupling constants. For the building blocks,
	\begin{gather}
	\chi_{+}=u^{\dagger}\chi u^{\dagger}+ u\chi^{\dagger}u,\\
	\chi={\rm diag}\left(m^{2}_{\pi}, m^{2}_{\pi}, -2m^{2}_{K}+3m^{2}_{\eta}\right).
	\end{gather}
	$\left\langle \chi_{+}\right\rangle$ is the trace of $\chi_{+}$, and $\tilde{\chi}_{+}=\chi_{+}-\frac{1}{3}\left\langle \chi_{+}\right\rangle$ is the traceless matrix.
	
	In HBChPT, the doubly bottom baryon field is expressed as the light component $H$ and the heavy component $h$,
	\begin{equation}
	\psi={\rm exp}\left(-im_{B}v\cdot x\right)\left(H+h\right),
	\end{equation}
	where $m_{B}$ are the masses of the doubly bottom baryons. The nonrelativistic Lagrangians were constructed in Refs. \cite{Sun: 2014aya, Qin: 2019hgk}. The lowest order nonrelativistic Lagrangian reads
	\begin{equation}
	\hat{\mathcal{L}}^{\left(1\right)}_{MB}=\bar{H}\left(iv\cdot D+g_{A}S\cdot u\right)H.
	\label{eq13}
	\end{equation}
	The $\mathcal{O}\left(p^{2}\right)$ and $\mathcal{O}\left(p^{3}\right)$ nonrelativistic Lagrangians read
	\begin{gather}
	\begin{aligned}
	\hat{\mathcal{L}}^{\left(2\right)}_{MB}=&\bar{H}\left[c_{1}\left\langle\chi_{+}\right\rangle+c_{2}u^{2}+c_{3}\left\langle u^{2}\right\rangle-8m^{2}_{0}c_{4}\left(v\cdot u\right)^{2}-4m^{2}_{0}c_{5}\left\langle\left(v\cdot u\right)^{2}\right\rangle+2c_{6}\left[S_{\mu}, S_{\nu}\right]\left[u^{\mu}, u^{\nu}\right]+c_{7}\tilde{\chi}_{+}\right.\\
	&\left.+\frac{2}{m_{0}}\left(S\cdot D\right)^{2}-\frac{ig_{A}}{2m_{0}}\left\{S\cdot D, v\cdot u\right\}-\frac{g^{2}_{A}}{8m_{0}}\left(v\cdot u\right)^{2}\right]H,
	\label{eq14}\\
	\end{aligned}\\
	\hat{\mathcal{L}}^{\left(3\right)}_{MB}=\bar{H}\left(-2d_{1}S\cdot u\left\langle\chi_{+}\right\rangle-2d_{2}\left\{S\cdot u, \tilde{\chi}_{+}\right\}-2d_{3}\left\langle S\cdot u\tilde{\chi}_{+}\right\rangle-\frac{i}{m^{2}_{0}}S\cdot Dv\cdot DS\cdot D+\cdots\right)H.
	\label{eq15}
	\end{gather}
	\section{The mass formulas of the doubly bottom baryons}
	\label{sec3}
	In HBChPT, with the variables of the self-energy, $\eta=v\cdot p-m_{B}$ and $\xi=\left(p-m_{B}v\right)^{2}$, the full propagator of baryons reads
	\begin{equation}
	iG=\frac{i}{v\cdot p-m_{0}-\Sigma_{B}\left(\eta, \xi\right)}=\frac{iZ_{N}}{\eta-Z_{N}\tilde{\Sigma}_{B}\left(\eta, \xi\right)},
	\end{equation}
	where $\Sigma_{B}\left(\eta, \xi\right)=\Sigma_{B}\left(0, 0\right)+\eta\Sigma^{\prime}_{B}\left(0, 0\right)+\tilde{\Sigma}_{B}\left(\eta, \xi\right)$ is the self-energy, and $Z_{N}=\frac{1}{1-\Sigma^{\prime}_{B}\left(0, 0\right)}$ is the wave function renormalization constant. Thus, the masses of the doubly bottom baryons read
	\begin{equation}
	m_{B}=m_{0}+\Sigma_{B}\left(0, 0\right).
	\end{equation}
	
\begin{figure}[htbp]
		\centering
		\includegraphics[width=0.5\hsize]{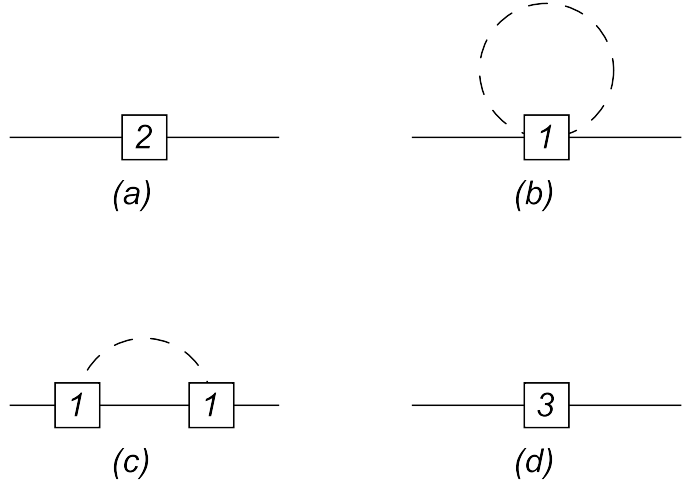}
		\caption{The Feynman diagrams contributing to the self-energies of the doubly bottom baryons up to $\mathcal{O}\left(p^{3}\right)$. The dashed and solid lines are the Goldstone bosons and the doubly bottom baryons. The numbers in the squares are the chiral orders of the vertices.}
		\label{fig1}
	\end{figure}

	We show the Feynman diagrams contributing to the self-energies of the doubly bottom baryons up to $\mathcal{O}\left(p^{3}\right)$ in Fig. \ref{fig1}. The vertice of in Fig. 1(a) is from the Lagrangian in Eq. (\ref{eq14}), the meson-baryon vertexs in Figs. 1(b) and 1(c) are from the Lagrangian in Eq. (\ref{eq13}), and the vertice in Fig. 1(d) is from the Lagrangian in Eq. (\ref{eq15}). In HBChPT, the chiral order of Feynman diagrams reads \cite{Ecker: 1994gg}
	\begin{equation}
	D_{\chi}=4N_{L}-2I_{M}-I_{B}+\sum_{n}nN_{n},
	\end{equation}
	where $N_{L}$ is the number of loops, $I_{M}$ and $I_{B}$ are the number of internal meson and baryon lines, and $N_{n}$ is the number of vertices from the $n$th order Lagrangians. Thus, $D_{\chi}=2$ in Fig. 1(a), and $D_{\chi}=3$ in Figs. 1(b)-1(d).
	
	After the calculations, Figs. 1(a) and 1(c) contribute to the $\mathcal{O}\left(p^{2}\right)$ tree-level masses and the $\mathcal{O}\left(p^{3}\right)$ loop-level masses, and Figs. 1(b) and 1(d) vanish. The masses of the doubly bottom baryons up to $\mathcal{O}\left(p^{3}\right)$ read
	\begin{equation}
	m_{B}=m_{0}-c_{1}\left\langle\chi_{+}\right\rangle-c_{7}\tilde{\chi}_{+}-\frac{g^{2}_{A}}{128\pi}\sum_{P}\frac{C_{PB}m^{3}_{P}}{F^{2}_{P}},
	\label{eq19}
	\end{equation}
	where the values of $C_{PB}$ are shown in Table \ref{table1}, and $m_{P}$ are the masses of the Goldstone bosons. We use the experimental values of $m_{P}$ \cite{ParticleDataGroup: 2020ssz}, $m_{\pi}=140{\rm MeV}$, $m_{K}=494{\rm MeV}$, $m_{\eta}=550{\rm MeV}$. After the combinations, Eq. (\ref{eq19}) reads
	\begin{gather}
	m_{\Xi_{bb}}=m_{0}-2c_{1}\left(2m^{2}_{\pi}-2m^{2}_{K}+3m^{2}_{\eta}\right)-\frac{2}{3}c_{7}\left(m^{2}_{\pi}+2m^{2}_{K}-3m^{2}_{\eta}\right)-\frac{g^{2}_{A}}{384\pi}\left(\frac{9m^{3}_{\pi}}{F^{2}_{\pi}}+\frac{6m^{3}_{K}}{F^{2}_{K}}+\frac{m^{3}_{\eta}}{F^{2}_{\eta}}\right),
	\label{eq20}\\
	m_{\Omega_{bb}}=m_{0}-2c_{1}\left(2m^{2}_{\pi}-2m^{2}_{K}+3m^{2}_{\eta}\right)+\frac{4}{3}c_{7}\left(m^{2}_{\pi}+2m^{2}_{K}-3m^{2}_{\eta}\right)-\frac{g^{2}_{A}}{96\pi}\left(\frac{3m^{3}_{K}}{F^{2}_{K}}+\frac{m^{3}_{\eta}}{F^{2}_{\eta}}\right),
	\label{eq21}
	\end{gather}
	\begin{table}[htbp]
		\centering
		\setlength{\tabcolsep}{2mm}{
			\begin{tabular}{cccc}
				\toprule[1pt]
				\toprule[1pt]
				~&$C_{\pi B}$&$C_{KB}$&$C_{\eta B}$\\
				\midrule[1pt]
				$\Xi_{bb}$&$3$&$2$&$\frac{1}{3}$\\
				$\Omega_{bb}$&$0$&$4$&$\frac{4}{3}$\\
				\bottomrule[1pt]
				\bottomrule[1pt]
		\end{tabular}}
		\caption{The values of $C_{PB}$.}
		\label{table1}
	\end{table}
	\section{Numerical results}
	\label{sec4}
	We need to determine the low-energy constants, $m_{0}$, $c_{1}$, $c_{7}$, and $g_{A}$ in Eqs. (\ref{eq20}) and (\ref{eq21}). Due to the limited experimental data, we have to perform the numerical processes by other theoretical information.
	
	For the doubly bottom baryons, $g_{A}$ was determined in the quark model \cite{Li: 2017cfz}, $g_{A}\left(bbq\right)=-0.50\left(5\right)$. Note that we add to the systematic error of $10\%$ for $g_{A}$ in the quark model.
	
	After the lattice QCD simulations, the authors performed the chiral extrapolations for the subtracted masses $E^{\left({\rm sub}\right)}_{X}$, rather than the full masses $E_{X}$ of the doubly heavy baryons in Ref. \cite{Brown: 2014ena},
	\begin{equation}
	E^{\left({\rm sub}\right)}_{X}=E_{X}-\frac{n_{c}}{2}\overline{E}_{c\overline{c}}-\frac{n_{b}}{2}\overline{E}_{b\overline{b}},
	\end{equation}
	where $n_{c}$ and $n_{b}$ are the number of $c$ and $b$ quarks in the doubly heavy baryons, and $\overline{E}_{c\overline{c}}$ and $\overline{E}_{b\overline{b}}$ are the spin-averaged charmonium and bottomonium masses. Then, the extrapolated values of $E^{\left({\rm sub}\right)}_{X}$ were added to the experimental values of $\frac{n_{c}}{2}\overline{E}_{c\overline{c}}+\frac{n_{b}}{2}\overline{E}_{b\overline{b}}$, $\overline{E}_{c\overline{c}}=3069{\rm MeV}$ and $\overline{E}_{b\overline{b}}=9445{\rm MeV}$  \cite{ParticleDataGroup: 2020ssz}, and the full masses of the doubly heavy baryons were obtained.
	
	We fit to three sets of lattice QCD data in Table \ref{table2} to determine $m_{0}$, $c_{1}$, and $c_{7}$. For convenience, the lattice QCD values of $E^{\left({\rm sub}\right)}_{X}$ were added to the experimental values of $\frac{n_{c}}{2}\overline{E}_{c\overline{c}}+\frac{n_{b}}{2}\overline{E}_{b\overline{b}}$. The $\pi$ and $\eta_{s}$ meson masses are used to set the $u/d$ and $s$ quark masses \cite{Davies: 2009tsa, HPQCD: 2011qwj}, and the $K$ and $\eta$ meson masses read
	\begin{gather}
	m_{K}=\sqrt{\frac{1}{2}\left(m^{2}_{\pi}+m^{2}_{\eta_{s}}\right)},\\
	m_{\eta}=\sqrt{\frac{1}{3}\left(m^{2}_{\pi}+2m^{2}_{\eta_{s}}\right)}.
	\end{gather}
	Then, we obtain the results of $m_{0}$, $c_{1}$, and $c_{7}$, and predict the masses of the doubly bottom baryons, as shown in Table \ref{table3}.
	
	For the charmed bottom baryons, we use the mass formulas of the doubly bottom baryons due to the heavy quark symmetry. We treat the heavy quarks as a diquark, including the symmetric spin diquark $\{cb\}$ and the antisymmetric spin diquark $[cb]$. $g_{A}$ was determined in the quark model \cite{Li: 2017cfz}, $g_{A}\left(\{cb\}q\right)=-0.50\left(5\right)$ and $g_{A}\left([cb]q\right)=1.51\left(15\right)$. With the same lattice QCD schemes, we obtain the results of $m_{0}$, $c_{1}$, and $c_{7}$, and predict the masses of the charmed bottom baryons, as shown in Table \ref{table3}.
	\begin{table}[htbp]
		\centering
		\setlength{\tabcolsep}{2mm}{
			\begin{tabular}{ccccccccc}
				\toprule[1pt]
				\toprule[1pt]
				~&$m_{\pi}$&$m_{\eta_{s}}$&$m_{\Xi_{bb}}$&$m_{\Omega_{bb}}$&$m_{\Xi_{\{cb\}}}$&$m_{\Omega_{\{cb\}}}$&$m_{\Xi_{[cb]}}$&$m_{\Omega_{[cb]}}$\\
				\midrule[1pt]
				I&$245$&$761$&$10227\left(13\right)$&-&$7016\left(14\right)$&-&$6969\left(13\right)$&-\\
				II&$270$&$761$&$10219\left(14\right)$&-&$7026\left(14\right)$&-&$6980\left(13\right)$&-\\
				III&$336$&$761$&$10224\left(13\right)$&$10297\left(13\right)$&$7027\left(13\right)$&$7104\left(13\right)$&$6982\left(13\right)$&$7065\left(13\right)$\\
				\bottomrule[1pt]
				\bottomrule[1pt]
		\end{tabular}}
		\caption{The lattice QCD data in Ref. \cite{Brown: 2014ena}. All the parameters are in ${\rm MeV}$.}
		\label{table2}
	\end{table}
	\begin{table}[htbp]
		\centering
		\setlength{\tabcolsep}{2mm}{
			\begin{tabular}{ccccccc}
				\toprule[1pt]
				\toprule[1pt]
				~&$m_{0}\left({\rm MeV}\right)$&$c_{1}\left({\rm GeV}^{-1}\right)$&$c_{7}\left({\rm GeV}^{-1}\right)$&$m_{\Xi_{QQ}}\left({\rm MeV}\right)$&$m_{\Omega_{QQ}}\left({\rm MeV}\right)$&$\chi^{2}_{\rm d.o.f.}$\\
				\midrule[1pt]
				$\Xi_{bb}/\Omega_{bb}$&$10264\left(53\right)$&$-0.015\left(36\right)$&$-0.104\left(20\right)$&$10235\left(63\right)$&$10299\left(64\right)$&$0.182$\\
				$\Xi_{\{cb\}}/\Omega_{\{cb\}}$&$7000\left(51\right)$&$-0.059\left(34\right)$&$-0.108\left(20\right)$&$7010\left(60\right)$&$7078\left(61\right)$&$0.161$\\
				$\Xi_{[cb]}/\Omega_{[cb]}$&$6854\left(48\right)$&$-0.326\left(32\right)$&$-0.323\left(20\right)$&$6930\left(63\right)$&$7017\left(84\right)$&$0.148$\\
				\bottomrule[1pt]
				\bottomrule[1pt]
		\end{tabular}}
		\caption{The results of $m_{0}$, $c_{1}$, and $c_{7}$ with the errors from the lattice QCD data, and the masses of the doubly bottom baryons and the charmed bottom baryons with the errors from $m_{0}$, $c_{1}$, $c_{7}$, and $g_{A}$.}
		\label{table3}
	\end{table}
	
\begin{figure}[htbp]
		\centering
		\includegraphics[width=1.0\hsize]{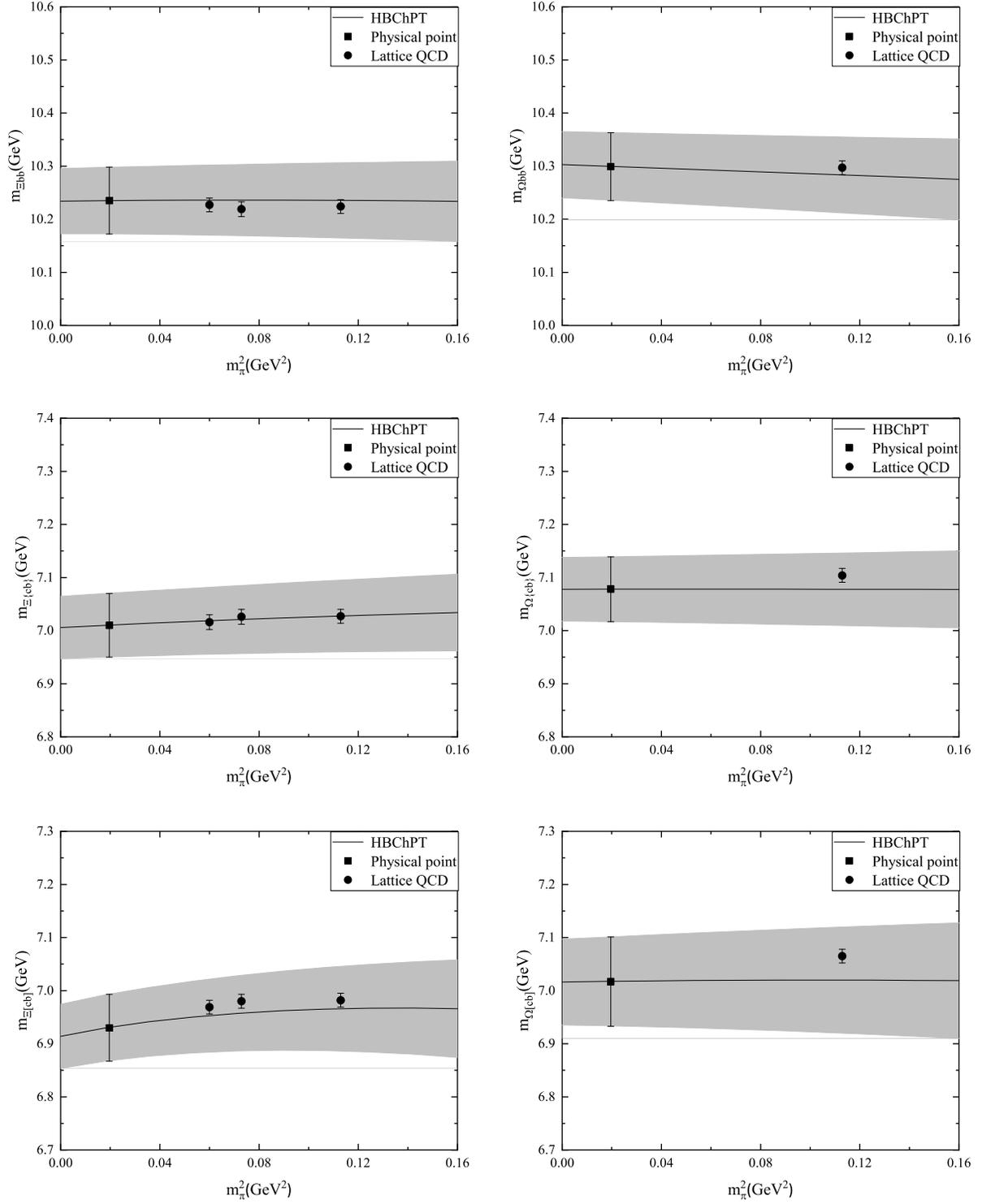}
		\caption{The masses of the doubly bottom baryons and the charmed bottom baryons as functions of $m^{2}_{\pi}$ with the error bands, and our results and the lattice QCD data with the error bars.}
		\label{fig2}
	\end{figure}

	We show each order contribution to the masses of the doubly bottom baryons and the charmed bottom baryons in Table \ref{table4}. We plot the masses of the doubly bottom baryons and the charmed bottom baryons as functions of $m^{2}_{\pi}$, and show our results and the lattice QCD data in Fig. \ref{fig2}.
	\begin{table}[htbp]
		\centering
		\setlength{\tabcolsep}{2mm}{
			\begin{tabular}{ccccc}
				\toprule[1pt]
				\toprule[1pt]
				~&$m_{0}$&$\mathcal{O}\left(p^{2}\right)$ tree&$\mathcal{O}\left(p^{3}\right)$ loop&$m_{B}$\\
				\midrule[1pt]
				$\Xi_{bb}$&$10264$&$-14$&$-15$&$10235$\\
				$\Omega_{bb}$&$10264$&$69$&$-34$&$10299$\\
				$\Xi_{\{cb\}}$&$7000$&$25$&$-15$&$7010$\\
				$\Omega_{\{cb\}}$&$7000$&$112$&$-34$&$7078$\\
				$\Xi_{[cb]}$&$6854$&$213$&$-137$&$6930$\\
				$\Omega_{[cb]}$&$6854$&$471$&$-308$&$7017$\\
				\bottomrule[1pt]
				\bottomrule[1pt]
		\end{tabular}}
		\caption{Each order contribution to the masses of the doubly bottom baryons and the charmed bottom baryons. All the parameters are in ${\rm MeV}$.}
		\label{table4}
	\end{table}
	\section{Summary}
	\label{sec5}
	We study the masses of the doubly bottom baryons and the charmed bottom baryons up to $\mathcal{O}\left(p^{3}\right)$ in HBChPT. For the low-energy constants, due to the limited experimental data, $g_{A}$ was determined in the quark model \cite{Li: 2017cfz}, and we have to perform the lattice QCD fits. For the doubly bottom baryons and the charmed bottom baryons, we fit to the lattice QCD data in Ref. \cite{Brown: 2014ena} to determine $m_{0}$, $c_{1}$, and $c_{7}$. We show the numerical results up to $\mathcal{O}\left(p^{3}\right)$, and our results are compared with other theoretical results in Table \ref{table5}.
	\begin{table}[htbp]
		\centering
		\setlength{\tabcolsep}{1mm}{
			\begin{tabular}{c|c|cccccc}
				\toprule[1pt]
				\toprule[1pt]
				~&Cases&$m_{\Xi_{bb}}$&$m_{\Omega_{bb}}$&$m_{\Xi_{\{cb\}}}$&$m_{\Omega_{\{cb\}}}$&$m_{\Xi_{[cb]}}$&$m_{\Omega_{[cb]}}$\\
				\midrule[1pt]
				Our results&~&$10235\left(63\right)$&$10299\left(64\right)$&$7010\left(60\right)$&$7078\left(61\right)$&$6930\left(63\right)$&$7017\left(84\right)$\\
				\midrule[1pt]
				\multirow{4}{*}{Quark models}&Ref. \cite{Weng: 2018mmf}&$10168.9\left(9.2\right)$&$10259.0\left(15.5\right)$&$6947.9\left(6.9\right)$&$7047.0\left(9.3\right)$&$6922.3\left(6.9\right)$&$7010.7\left(9.3\right)$\\
				~&Refs. \cite{Salehi: 2019kvb, Mohajery: 2018qhz}&$9716$&$10870$&$6628$&$7329$&$6628$&$7329$\\
				~&Refs. \cite{Karliner: 2014gca, Karliner: 2018hos}&$10162\left(12\right)$&$10208\left(18\right)$&$6933\left(12\right)$&$6984\left(19\right)$&$6914\left(13\right)$&$6968\left(19\right)$\\
				~&Ref. \cite{Shi: 2019tji}&$10210$&$10319$&-&-&-&-\\
				\midrule[1pt]
				\multirow{5}{*}{Lattice QCD}&Ref. \cite{Brown: 2014ena}&$10143\left(30\right)\left(23\right)$&$10273\left(27\right)\left(20\right)$&$6959\left(36\right)\left(28\right)$&$7032\left(28\right)\left(20\right)$&$6943\left(33\right)\left(28\right)$&$6998\left(27\right)\left(20\right)$\\
				~&Ref. \cite{Lewis: 2008fu}&$10127\left(13\right)\left(^{12}_{26}\right)$&$10225\left(9\right)\left(^{12}_{13}\right)$&-&-&-&-\\
				~&Ref. \cite{Burch: 2015pka}&$10267\left(44\right)\left(61\right)$&$10356\left(34\right)\left(68\right)$&$6937\left(105\right)\left(46\right)$&$7077\left(109\right)\left(47\right)$&$6887\left(103\right)\left(46\right)$&$7039\left(108\right)\left(47\right)$\\
				~&Ref. \cite{Mathur: 2018epb}&-&-&$6966\left(23\right)\left(14\right)$&$7045\left(16\right)\left(13\right)$&$6945\left(22\right)\left(14\right)$&$6994\left(15\right)\left(13\right)$\\
				~&Ref. \cite{Mohanta: 2019mxo}&$10091\left(17\right)$&$10190\left(17\right)$&$6843\left(19\right)$&$6946\left(17\right)$&$6787\left(12\right)$&$6893\left(16\right)$\\
				\bottomrule[1pt]
				\bottomrule[1pt]
		\end{tabular}}
		\caption{Our results and other theoretical results for the masses of the doubly bottom baryons and the charmed bottom baryons, including quark models \cite{Weng: 2018mmf, Mohajery: 2018qhz, Salehi: 2019kvb, Karliner: 2014gca, Karliner: 2018hos, Shi: 2019tji} and lattice QCD \cite{Brown: 2014ena, Lewis: 2008fu, Burch: 2015pka, Mathur: 2018epb, Mohanta: 2019mxo}. All the parameters are in ${\rm MeV}$.}
		\label{table5}
	\end{table}
	
	The properties of the doubly heavy baryons are still worth exploring, which can help us to understand the mechanism of the hadron spectrum. With the discovery of the $\Xi^{++}_{cc}$ baryon, we hope for more researches on the doubly heavy baryons. Our numerical results may be useful for future experiments.
	\section*{Acknowledgments}
	This project is supported by the National Natural Science Foundation of China under
	Grants No. 11905171 and No. 12047502. This work is also supported by the Double First-class University Construction Project of Northwest University.
	
\end{document}